\documentclass[
  10pt,
  reprint,
  showpacs,preprintnumbers,
  amsmath,amssymb,
  longbibliography,
  aps,
  prd,
]{revtex4-1}

\makeindex 
\usepackage{graphicx}
\usepackage[colorlinks=true, urlcolor=blue, linkcolor=blue, citecolor=blue]{hyperref}

\usepackage{definition}
\renewcommand{\BibitemShut}[1]{}
\begin{document}

\title{ Effective confining potentials for QCD}

\author{ Arkadiusz P. Trawi\'nski and Stanis{\l}aw D. G{\l}azek}
\affiliation{ Institute of Theoretical Physics,
     Faculty of Physics, 
     University of Warsaw,
     Ho\.za 69,
     00-681 Warsaw, Poland }

\author{Stanley J. Brodsky}
\affiliation{
  SLAC National Accelerator Laboratory,
  Stanford University,
  Stanford, California 94309,USA}

\author{ Guy F. de T\'eramond}
\affiliation{Universidad de Costa Rica, 1000 San Jos\'e, Costa Rica} 

\author{ Hans G\"unter Dosch}
\affiliation {Institut f\"ur Theoretische Physik, Philosophenweg 16, Heidelberg, Germany}


\date{ \today{} }

\preprint{SLAC-PUB-15924}
\preprint{IFT/14/01}

\begin{abstract}
We observe that the linear potential used as a leading approximation for describing
color confinement in the instant form of dynamics corresponds to a quadratic confining
potential in the front form of dynamics. In particular, the instant-form potentials
obtained from lattice gauge theory and string models of hadrons agree with the potentials
determined from models using front-form dynamics and light-front holography, not only
in their shape, but also in their numerical strength.
\end{abstract}

\maketitle

\section{Introduction}
A key question in QCD is to understand the nonperturbative
dynamics underlying the confinement of quarks and gluons~\cite{GellMann:1962xb,
GellMann:1964nj} from first principles. Various approaches to nonperturbative QCD
such as  lattice gauge theory, AdS/QCD and string theory appear to describe color
confinement in very different ways. In this letter we will show that despite their
apparent analytic differences, these approaches have essential elements in common
if one takes into account the fact that the shape of the confinement potential 
depends on the form of dynamics; e.g., instant form (IF) versus the front 
form (FF)~\cite{Dirac:1949cp}, the latter called also the 
light-front (LF) dynamics in the literature.

Nonrelativistic analyses such as heavy quark effective theory are based on the usual 
IF dynamics where the Hamiltonian is the usual-time evolution operator.  
Relativistic bound-state problems such as confinement of light quarks are usually formulated 
in the FF Hamiltonian dynamic framework, since it provides a rigorous frame-independent
formalism. In this case, the LF Hamiltonian is the time-evolution operator
$H_{LF} = i{\partial\over\partial \tau}$ where $\tau = (ct+z)/c$ 
is the time along the light front.

It is important to note that the form of the effective potential in each
formalism depends on the form of the dynamics which is utilized.
In this paper we will compare the physical descriptions, their effective
potentials, and the mass scales controlling quark confinement obtained
from lattice, string theory, and the FF approach based on
LF holography. An essential observation is that a linear confining
potential in the IF of dynamics agrees with a quadratic confining potential
in the FF of dynamics at leading approximation. One thus obtains
a common element of quantum-mechanical effective theories which 
incorporates color confinement, relativity, and essential spectroscopic
and dynamical features of hadron physics. 

An important tool will be the Wentzel-Kramers-Brillouin (WKB)~\cite{WKB}
formalism which allows one to relate the maximum distance of separation 
between quarks within a meson as predicted by each model. We find that 
this parameter appears to be universal even among different forms of 
dynamics. It thus provides a universal point of focus for describing the
same phenomenon of color confinement in different approaches. 

We begin by recalling that the IF of the non-relativistic Schr\"odinger
operator for a system made of two strongly interacting particles of identical
mass $m$ and with momenta $\vec p_q = \vec p$ and $\vec p_{\bar q} = -\vec p$,
such as $J/\psi$, $\Upsilon$ or other mesons, is
\begin{align}
\label{eq:IF_equation}
  \cM 
& =
  2\, m + {\v p^2 \over m} + V_\text{eff} \,,
\end{align}
where $2 m + \frac{ \vec p\,^2}{m}$ originates from $2\sqrt{m^2+ \vec p\,^2}$.
The eigenvalue of $\cM$ is the mass of the system. In contrast,
the FF formulation of the theory of interacting particles is
applicable to non-relativistic as well as relativistic constituents.
It leads to an effective eigenvalue equation for the mass squared operator
\begin{align}
\label{eq:FF_equation}
  \cM^2
& =
  {k_\perp^{\,2} + m^2 \over x(1-x) } + U_\text{eff} \,,
\end{align}
instead of $\cM$ in the IF of dynamics, Eq.~(\ref{eq:IF_equation}). 
The boost-invariant FF variables $x$ and $1-x$ are ratios of longitudinal
FF momenta $p^+_{q} = p^t_{q} + p^z_{q}$ and $p^+_{\bar q} = p^t_{\bar q} + p^z_{\bar q}$
of the constituents to the longitudinal FF momentum of the meson, $P^+ = P^t + P^z$.
The term $k_\perp^{\,2} + m^2 \over x(1-x) $ is the LF kinetic energy as well as
the invariant mass squared $s = (p_q + p_{\bar q})^2$ of the $q \bar q$ pair.

It will be convenient to define a relative three-vector momentum operator $\vec p$
(in the constituent rest frame~\cite{Danielewicz:1978mk,Karmanov:1979if,Glazek:1983ba}),
so that
\begin{align}
\label{eq:NR_FF}
{\cal M}^2
={ k^2_\perp +m^2 \over x(1-x)}  + U_\text{eff}
\equiv 4\,m^2 + 4\,{\vec p }^{\,2} + U_\text{eff}.
\end{align}
We identify $p^2_\perp = {k_\perp^2 \over 4 x(1-x)}$
and $4\,m^2 + 4\, p_3^2 = {m^2 \over  x(1-x) }$, so
$p_3 =  {m \over \sqrt{x(1-x)} }\big(x - \frac{1}{2}\big)$
has an infinite range and is proportional to $m$. The conjugate
variables are $r_\perp= i \frac{\partial}{\partial p_\perp}$
and $r_3 = i \frac{\partial }{ \partial p_3}$. Early discussions
of models of mesons as two-body systems in the FF dynamics, as 
alternative to the IF, especially in the infinite momentum frame, 
can be found in Refs.~\cite{Melosh:1974cu,Leutwyler:1977vz, Leutwyler:1977pv,
Leutwyler:1977cs, Leutwyler:1977vy,Bakker:1979eg}.

The central problem then becomes the derivation of the
effective interactions $V_\text{eff}$ and $U_\text{eff}$.
We observe that nearly all considerations in the IF of
the Hamiltonian dynamics lead to the conclusion that
the potential between a quark and antiquark at large
distances should be linear. This article points out
that the linear IF potential $V_\text{eff}$ implies
a quadratic FF potential $U_\text{eff}$ at large $q \bar q$
separation, as implied by Eqs.(\ref{eq:IF_equation})
and~(\ref{eq:NR_FF}),
\begin{align}
\label{eq:U_eff&V_eff}
U_\text{eff} &=  V^2_\text{eff} + 2 \sqrt{\v p^2 + m^2} \,  V_\text{eff}
               + 2 \,  V_\text{eff} \sqrt{\v p^2 + m^2} \,.
\end{align}
At large distances, near turning points where kinetic
energy is minimal, the potential term $V^2_\text{eff}$
dominates the right-hand side. 
Thus, for a linear IF potential $V_\text{eff}$,
the FF potential $U_\text{eff}$ is quadratic. 
Such FF harmonic oscillator potential predicts  
linear Regge trajectories~\cite{Regge:1959mz,Regge:1960zc}
in the hadron mass square for small quark masses.

In the next sections, various models defined in different forms
of dynamics will be discussed. In the last section we will
compare the models in terms of their WKB parameter.

\vspace{-1.0em}
\section{Instant-form approaches}
\vspace{-1.0em}

The main contemporary tool for studying mesons in the IF 
of dynamics is lattice QCD, originally formulated in 
Ref.~\cite{Wilson:1974sk}. One obtains numerical results 
for hadron properties from calculating their Euclidean 
propagators~\cite{Hagler:2009ni}. The underlying dynamics 
can be studied in terms of a potential by calculating 
the Wilson loops, where quarks are represented by static 
color sources~\cite{Kogut:1974, Bardeen:1976tm}. We focus
first on methods which allow one to compute the shape and mass 
scale of the non-relativistic potential which confines
pairs of infinitely heavy quarks~\cite{Appelquist:1974zd,
Appelquist:1974yr, Eichten:1974af}. The lattice approach
is closely related to the string picture for hadrons 
(see below).

\vspace{-1.5em}
\subsection{Specific lattice-potential results}
\vspace{-1.0em}

The static potential obtained in the quenched approximation of the lattice QCD
can be parameterized in the form of the Cornell potential~\cite{Eichten:1978tg};
\emph{i.e.} (up to a constant term)
\begin{align}
\label{eq:V_lattice}
  V_\text{eff}^{(lattice)} (r)
& =
  - {A \over r} + \sigma r \,,
\end{align}
where $r$ denotes the distance between infinitely heavy (static) quark
and antiquark and $\sigma$ is called string tension. The string tension
due to the gluonic fields connecting static color sources does not
include the pair creation mechanism that breaks the string; there is
thus no direct relation of the two-body effective potential for QCD
to this aspect of the string tension

The progress made in simulations of QCD on the lattice allows one to 
calculate coefficients $A$ and $\sigma$ also for quarks with finite 
masses. For instance, one of the most recent 
analyses of charmonium~\cite{Kawanai:2011xb,Kawanai:2011jt} found the 
square root of string tension of magnitude $\sqrt \sigma = 394(7)$ MeV, 
associated with the quark mass $1.74(3)$ GeV. Despite the fact that our 
discussion concerns quarks with the phenomenological values of masses 
which may be different from the charmonium result of $1.74$ GeV, the 
value of $\sqrt \sigma = 394$ MeV appears appropriate for our purpose 
of estimating the behavior of the quark-antiquark effective potential
in the configuration where the potential dominates the meson energy. 
However, it should be mentioned that in the case of static sources the 
value of $\sqrt \sigma \sim 460$ MeV is obtained~\cite{jldg, Aoki:2008sm, 
Koma:2012bc}. Lattice estimates for the universal quark-antiquark
potentials should be based on calculations for quarks with finite effective
mass parameters.

\vspace{-1.5em}
\subsection{Classical string model}
\vspace{-1.0em}

An effective description of quark confinement in mesons is the string
model for hadrons, where color-electric fields between two static color sources
are squeezed into a thin, effectively one-dimensional, flux tube or
vortex~\cite{Abrikosov:1956sx, Nielsen:1973cs, 'tHooft:1973jz, Migdal:1984gj}.
The string picture of confinement can be considered~\cite{Kogut:1974}
as the strong coupling limit of the IF Hamiltonian formulation of lattice QCD.

One can study the spectra of multi-dimension string models~\cite{Luscher:1980iy,
Luscher:1980ac, Luscher:1980fr,Andreev:2006ct} such as strings described by
the Nambu-Goto action~\cite{Goto:1971ce,Nambu:1974zg}. This approach yields
a string with a constant energy density per unit length and a static potential
which rises linearly as a function of the string length $r$. In the 4-dimensional
space-time, the quark-antiquark potential is thus given (up to a constant 
term) by~\cite{Luscher:1980fr, Arvis:1983fp}
\begin{align}
\label{eq:V_string}
  V_\text{eff}^{(string)} (r)
& =
  \sigma r \sqrt{1-{\pi \over 6\,\sigma\, r^2}} \,.
\end{align}
From this, one can calculate the dependence of the meson spectrum on the internal 
angular momentum. By comparing with the empirical Regge trajectories, one finds 
a slope parameter  470 MeV $< \sqrt\sigma <$ 480 MeV for pseudo-scalars ($\pi$ 
and $K$), while for other mesons the value of $\sqrt{\sigma}$ varies between $424$ 
and $437$ MeV~\cite{Bali:2000gf}. The string description applies for distances 
$ r \gg r_c = \sqrt{\pi / (6 \sigma)}$, and $r_c \simeq 0.33$ fm for $\sqrt{\sigma} 
\simeq 430$ MeV. 
Most of the above results point toward the value about 
430 MeV with an ambiguity on the order of 7 MeV. A 
review of the lattice and the string theories can be 
found in Refs.~\cite{Bali:2000gf, Bazavov:2009bb}.

\vspace{-1.5em}
\subsection{Stochastic vacuum model}
\vspace{-1.0em}
In the stochastic vacuum model (SVM), string formation
is a property of the gauge-invariant gluon field-strength
correlator, which can be obtained by lattice simulations.
It thus connects the lattice with the hadronic string
picture~\cite{Dosch:1988ha, DiGiacomo:2000va}.

The SVM~\cite{Dosch:1988ha} starts with the assumption that
the nonperturbative (long-distance) part of the functional
integral over the gluon field can be approximated by a Gaussian
integration. Wilson loops can be evaluated easily and are
determined by the gauge-invariant correlator of the gluon fields;
for large loops one derives an area law signifying linear confinement.
The resulting nonrelativistic potential begins quadratically and becomes
linear at distances comparable to the correlation length of the gluon
field. The confinement mechanism is due to the formation of a color-electric
string between the quark and antiquark~\cite{DiGiacomo:2000va}.
The string tension is given by~\cite{Dosch:1988ha,DiGiacomo:2000va}
\begin{align}
\label{eq:sigma_svm}
\sigma = \frac{\pi}{48 \, N_c} \int_0^\infty d z^2\, D(z^2) \,,
\end{align}
where $D(z^2)$  is the scalar part of the gauge invariant colour  field
correlator $\langle\, G_{\mu \nu}(z) \, \Phi(z) \, G_{\rho \sigma}(0)\,\rangle$
and $\Phi(z)$ is the colour transporter from point 0 to $z$.
$D(z^2)$ can be calculated on the lattice using
the cooling method~\cite{D'Elia:1997ne}. Using the numerical results of this
lattice simulation~\cite{D'Elia:1997ne}  one obtains for the string tension
$\sqrt{\sigma} = 410(11)$MeV.

\vspace{-1.0em}
\section{Front-form approaches}
\vspace{-1.0em}
The lattice gauge theories are not effectively formulated using the FF
of dynamics because of difficulties with understanding what to do in
the Minkowski space. High-energy experiments require an efficient IF
description in the infinite momentum frame or, in a frame-independent
way, or description using the FF. Since there is no efficient lattice description
that could be used, one turns to the FF Hamiltonian methods. 

The derivation of the FF QCD Hamiltonian eigenvalue equation 
that accounts for dynamical effects of all virtual quarks 
and gluons present in the Fock-space expansion of a hadron 
state, requires a suitable renormalization group procedure.
We focus on the procedure called the similarity renormalizarion
group procedure~\cite{Glazek:1993rc, Glazek:1994qc, Wilson:1994fk},
and to its successors, especially the renormalization group 
procedure for effective particles (RGPEP, see below).

The potential $U_\text{eff}$ in the FF effective Hamiltonian,
Eq.~(\ref{eq:FF_equation}), can be found by applying the Ehrenfest
principle~\cite{Ehrenfest} to quantum field theory~\cite{Glazek:2013jba},
in the sense of calculating expectation values which average
quantities of interest over all Fock sectors and all 
effective constituents in them, except for the constituent that
is struck by an external probe, called the active one.
In every Fock sector, the active constituent moves in an effective
potential generated by the remaining constituents, called spectators.
Following this line of reasoning, the resulting potential 
describes the motion of an active constituent around the minimum of
its potential energy.
Such a potential is expected to be quadratic, $U_\text{eff} (r) \sim r^2 $,
as every regular function around its minimum is. Both
the Ehrenfest equation and the quadratic potential agree with
the requirement of rotational symmetry because all Fock sectors
in the bound-state dynamics are included,~\emph{cf.}~\cite{Perry:1994kp,
Brisudova:1996vw, Brisudova:1995hv}; \emph{i.e.} multiplets of
the spectrum have the mass degeneracy required by the rotational
symmetry in 3-dimensions. The quadratic form of the FF Ehrenfest
potential around its minimum agrees well with the large-$r$ 
result that $U_\text{eff} (r) \sim r^2 $ when $V_\text{eff} (r) \sim r $,
and with results of the LF holography. This will be explained after
we discuss the LF holography.

\vspace{-1.5em}
\subsection{LF holography}
\vspace{-1.0em}
One can write the FF equation of motion for mesons
in the form of a single-variable relativisitic eigenvalue equation
analogous to the non-relativistic quark-antiquark radial Schr\"odinger
equation~\cite{deTeramond:2005su}. The same equation for massless
quarks arises from the LF holographic mapping~\cite{deTeramond:2008ht,
deTeramond:2013it} of the soft-wall model modification of AdS$_5$
space~\cite{Karch:2006pv} with any dilaton profile which breaks
the maximal symmetry of AdS$_5$ space. Thus one arrives at a meson
equation of motion for zero quark mass, where the fifth-dimension
variable $z$ in AdS$_5$ becomes identified with the boost-invariant
transverse $q \bar q$ separation variable $\zeta$. One has
$\zeta^2 = {1\over4}\,r_\perp^2 = x(1-x) b^2_\perp$, where
$b_\perp = i{\partial \over \partial k_\perp}$ is the transverse
distance between the two constituents~\cite{Brodsky:2007hb}.
The resulting single-variable relativistic equation of motion
includes a harmonic oscillator potential 
\begin{align}
\label{eq:U_LF}
  U_\text{eff}^{(LF)}(\zeta)
& =
  \kappa^4 \zeta^2 + 2 \kappa^2(J-1) \,,
\end{align}
where $J$ is the total angular momentum of the $q\bar q$ meson.
The LF-holography is inspired by Maldacena
conjecture~\cite{Maldacena:1997re}; it does not require that
the number of colors is large.

It has been shown that the harmonic oscillator form
of the FF potential arises uniquely when one extends
the formalism of de Alfaro, Fubini and Furlan~\cite{deAlfaro:1976je}
to the FF Hamiltonian theory~\cite{Brodsky:2013ar}.
The action of the effective one-dimensional quantum
field theory remains conformally invariant, which reflects
the underlying conformal invariance of the classical QCD
chiral Lagrangian. The constant term $2 \kappa^2(J-1)$
is derived from spin-$J$ representations of dynamics
in AdS space~\cite{deTeramond:2013it}.

The mass parameter $\kappa$ is determined outside
of QCD from a single observable, such as the pion decay
constant. One finds consistency with hadron spectroscopy
for $\kappa$ between 540 and 590 MeV.

It is natural to replace $\kappa^4\,\zeta^2 = {1\over4}\,\kappa^4 r_\perp^2$
in Eq.~(\ref{eq:U_LF}) by ${1\over4}\,\kappa^4 (r_\perp^2 + r^2_3)$
in the case of massive quarks~\cite{Glazek:2011vg}, where
the quark masses are input parameters. Then $U_\text{eff}^{(LF)}$
becomes a 3-dimensional oscillator potential. The corresponding wave
function matches phenomenology, \emph{e.g.} see Ref.~\cite{Brodsky:1980vj}.
Thus, excitations in the transverse plane are paired with
excitations in the 3-direction, and 3-dimensional rotation
symmetry is restored in the massive case. This change
also establishes connection with the 3-dimensional Eq.~(\ref{eq:NR_FF}),
and it does not require any change of the value of $\kappa$.
The same universal value of $\kappa$ is also obtained 
when short-range spin-dependent interactions 
are included~\cite{Branz:2010ub,Gutsche:2012ez}. 
It would be interesting to extend these
results retaining 3-dimensional rotation symmetry.

\vspace{-1.5em}
\subsection{Gluon condensate model embedded\\ in the RGPEP}
\vspace{-1.0em}
A framework based on the RGPEP~\cite{Glazek:2011vg} allows
one to develop a relativistic quark model inspired
by~\cite{Glazek:1987ic}, where the effective particle masses
are different from zero and can be set as the input parameters. 
The FF potential is quadratic as a function of a 
3-dimensional quark-antiquark distance~$r$, 
\begin{align}
\label{eq:U_RGPEP}
  U_\text{eff}^{(RGPEP)} (r)
& =
  \left({\pi \over 3} \varphi_\text{glue}\right)^2 r^2 \,,
\end{align}
where $\varphi_\text{glue}^2$ represents the gluon 
condensate inside hadrons.
In the operator product expansion~\cite{Wilson:1969zs} the expectation value 
corresponding to gluon condensate can also refer to matrix elements inside 
hadrons rather than the vacuum~\cite{Casher:1974xd,Maris:1997tm,Maris:1997hd,
Brodsky:2008be, Brodsky:2008xm,Brodsky:2010xf,Chang:2011mu}.

The original value of $\varphi_\text{glue}^2 = 0.012\ \text{GeV}^4$
obtained by Shifman, Vainshtein and Zakharov~\cite{Shifman:1978bx}
has been updated by Narison~\cite{Narison:1995tw,Narison:2011xe},
which in the case of in-hadron condensate implies (see Eq. (20)
in Ref.~\cite{Narison:2011xe})
\begin{align}
  \varphi_\text{glue}^2 & = {1\over\pi}{\bra{G}\, {\alpha_s} \, G^{\mu\nu c}G_{\mu\nu}^c \ket{G} \over \braket{G}{G} }
             \simeq \ 0.022(4)\ \text{GeV}^4 \,,
\end{align}
where $\alpha_s$ is the QCD coupling constant, and $\ket{G}$
represents the gluons condensed inside a meson.

\vspace{-1.5em}
\section{Discussion}
\vspace{-1.0em}
In the IF models, the confinement potential increases 
linearly at large distances between static quarks as 
exemplified in Eqs.~(\ref{eq:V_lattice}-\ref{eq:sigma_svm}). 
Other terms contribute at small distances. The eigenvalues
of the IF Hamiltonian are the energies of the hadrons.
In  contrast, the  eigenvalues of the Hamiltonian in
the frame-independent FF of  dynamics  is  quadratic
in the hadron  mass  $M$: Eqs.~(\ref{eq:U_LF}) 
and~(\ref{eq:U_RGPEP}). Note that $M^2 = (2m + \epsilon)^2
= 4m^2 + 4m \epsilon + \epsilon^2$ where $\epsilon$ is 
the binding energy. It is essential to retain the 
$\epsilon^2$-term in the FF eigenvalue equation, 
Eq.~(\ref{eq:FF_equation}), since the $\epsilon^2$-term 
contributes to the FF potential $U_\text{eff}$ even if 
$m$ is large, see Eq.~(\ref{eq:NR_FF}). 
Thus, if the IF potential is linear~\cite{Glazek:2003ky,
Glazek:2013jba}, then the FF potential in the 
non-relativistic limit at a large distance between 
quarks should be quadratic. This can be seen straightforwardly 
in the cases where the mass of constituents $m$ tends to zero.

\begin{figure}[!ht]
  \includegraphics[width=1.0\linewidth]{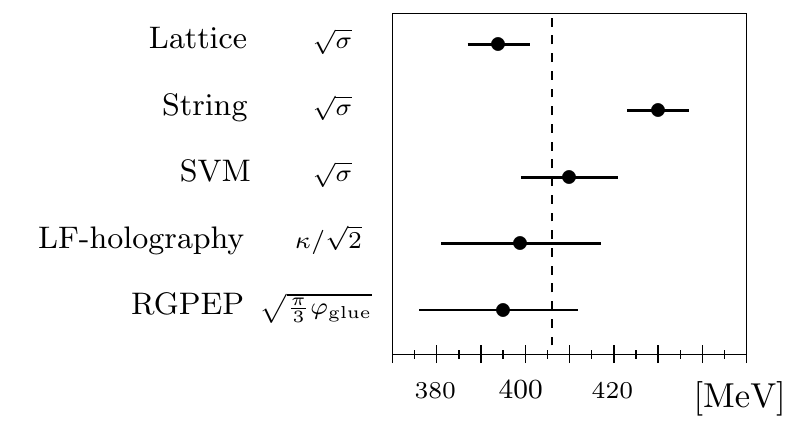}
  \caption[]
  {\label{fig:compere}
Phenomenological results for the coefficient of $r_\text{max}$ 
obtained using the WKB method (see the text). We compare the 
coefficients obtained from the lattice approach $\sqrt \sigma 
= 394(7)$ MeV, string theory $\sqrt{\sigma} = 430(7)$ MeV,
the stochastic vacuum model $\sqrt{\sigma} = 410(11)$ MeV,
the LF holography approach $\kappa/\sqrt{2} = 381 \div 417$ MeV,
and the in-hadron gluon condensate in the RGPEP approach 
$\sqrt{\pi \varphi_\text{glue}/3} = 395(18)$ MeV.
The dashed line is the average of these values.
}
\end{figure}

In order to compare different descriptions of confinement 
we can adopt the WKB method. It defines the turning point 
$r_\text{max}$ where the kinetic energy is completely turned 
into potential energy. One obtains $M = 2m + \sigma \, 
r_\text{max}$ in the lattice, string and SVM approach,
$M^2 = 4m^2 + ({\pi \over 3}\, \varphi_\text{glue})^2 \, 
r^2_\text{max}$ in the RGPEP approach, and $M^2 = 4m^2 + 
{1\over 4}\,\kappa^4 \, r^2_\text{max}$ in the LF-holography approach.
The last factor ${1\over 4}$ comes from the fact that $x = {1\over 2}$
at the WKB turning point, where $p_\perp$ and $p_3$ both vanish.

Figure~\ref{fig:compere} compares the phenomenological results 
for the coefficient of $r_\text{max}$. The values for the 
effective confinement scales derived from the WKB analysis 
in each model discussed above are sufficiently close to each 
other that one can argue that the various confinement models 
describe the same effective two-body system in the IF and in 
the FF of dynamics. There are different scales of energy in QCD, 
determined by quantities such as masses of quarks, $\Lambda_{QCD}$, 
both possibly multiplied by some powers of $\alpha_{QCD}$. 
Nevertheless, the values of parameters quoted here are of the 
same order. Finally, we wish to stress that the linear confining 
potential of the IF of dynamics is consistent with the quadratic 
confining potential in the FF of dynamics. 

\begin{acknowledgments}
APT wants to express his gratitude for the
hospitality extended to him at SLAC where this article was written.
This work was supported by the Polish-U.S. Fulbright Commission
and the Foundation for Polish Science International Ph.D. Projects Programme,
co-financed by the EU European Regional Development Fund.
\end{acknowledgments}

\bibliography{bibliography}

\end{document}